\documentclass[onecolumn,showpacs,showkeys,preprintnumbers,amsmath,amssymb]{revtex4}

\usepackage{setspace}
\usepackage{graphicx}
\usepackage{dcolumn}
\usepackage{bm}
\usepackage{epsfig}
\begin{document}

\title[]{Plasma: the genesis of the word}
\author{Mario J. Pinheiro}
\address{Department of Physics and Center for Plasma Physics,\&
Instituto Superior T\'{e}cnico, Av. Rovisco Pais, \& 1049-001
Lisboa, Portugal} \email{mpinheiro@ist.utl.pt}


\date{\today}%
\begin{abstract}
The historical roots of the word plasma are recalled. It is
suggested that possibly at the beginning of the researches in
low-pressure gas discharges the driving motivation was related to
the psychic phenomena investigated by Sir William Crookes.
\end{abstract}
\maketitle

Michel Foucault wrote that objects and words intercross deeply, and
Nature only offer herself to our understanding through words;
without the right words, Nature will subsume in a fading
light~\cite{Foucault}. This reasoning pushed us to recall how the
"fourth state of matter" was denominated. We believe that
understanding the genealogy of the word will throw some light to its
roots, the content and philosophy of this science.

In fact, the word {\it plasma} comes from the Greek ($\pi \lambda
\alpha \sigma \mu \alpha$) and means "to mold", "to shape". Jan
Evangelista Purkyn\v{e} (1787-1869), a notable Czech medical
scientist (best known for his discovery of Purkinje cells), used for
the first time this word to designate the reminescent translucid
liquid that remains after completely removing the blood from all the
corpuscules (he also introduced the word protoplasm for the
substance inside the cell).

Faraday pointed out that matter can be classified in four states:
solid, liquid, gas, and radiant (in his own words). The research on
this new form of matter was initiated by Heinrich Gei$\beta$ler
(1814-1879), a glassblower, who invented the sealed glass tube (now
called Gei$\beta$ler tube). Julius Pl\"{u}cker (1801-1868), while
professor at Bonn, published his results about the action of the
magnet on the electric discharge in rarefied gases. With his pupil
Johann Wilhelm Hittorf (1824-1914) he made many important
discoveries in the spectroscopy of gases~\cite{Hittorf,Tomaschek}.

Later on Sir William Crookes took again the term "radiant matter"
coined by Faraday, to designate this new "form" of matter. In his
own words:
\begin{quote}
«So distinct are these phenomena from anything which occurs in air
or gas at ordinary tension, that we are led to assume that we are
here brought face to face with matter in a fourth state or
condition, a condition as far removed from the state of gas as a gas
is from a liquid.»~\cite{Crookes_1}.
\end{quote}
The invention of the radiometer by Crookes lead him to the study of
rarefied gases and, in particular, to study electric discharges in
rarefied gases. This new line of research was open by the discovery
of cathode rays by Wilhelm Hittorf in 1869. According to Crookes
cathode rays were a kind of matter in the ultra-gaseous state,
emitted by the cathode and had the dimensions of molecules. They
were invisible but could be turned perceptible by the
phosphorescence that they induce on the walls of the glass tube or
by the shadows left behind by objects placed along their paths.

The researches done by Pl\"{u}cker, Hittorf and Crookes constitute
the first fundamental observations of new phenomena that led to the
new discipline of glow discharge plasma, dedicated to the study of a
very complex non-equilibrium plasma. A voltage were applied across
two electrodes placed inside in an optically transparent vacuum
envelope (with a pressure of the order of a few torr), usually made
of pyrex glass or quartz. The region near the cathode was later on
called the Hittorf-Crookes dark space; dark because the electrons
density is not enough to ionize the gas.

But experiments done in 1881 by J. J. Thomson have shown later that
this "radiant matter" were in fact constituted by electrified
corpuscules that were later named electrons (a term coined by
Johnstone Stoney~\cite{Stoney_1}), the separate units of
electricity, and that theory was abandoned.

It is quite interesting to notice that some esoteric notions used by
Crookes were possibly at the historical root of the word plasma. The
same pre-logical formulations were present at the genesis of
Newton's mechanics, endowing the gravitic force with esoteric
properties:
\begin{quote}
    This most beautiful system of the sun, planets, and comets,
    could only proceed from the counsel and dominion of an
    intelligent and power Being."~\cite{Newton_1}.
\end{quote}
But the word plasma may possibly had a previous meaning related to
phantoms, ghosts and others interests pursued on by Sir William
Crookes, and that ultimately led some people to denounce him as a
not honest, or lucid, scientist. Let's cite just a small paragraph
of his 1883 Bakerian Lecture given to the Royal Society of London:
\begin{quote}
    No Will-o'-the-Wisp ever led the unwary traveller into so many
    pitfalls and sloughs of despond as the hunt for this phantom
    band had entrapped me."~\cite{Crookes_2}
\end{quote}

Of course, we know that his open-mindedness lead him to investigate
psychic phenomena with several famous spiritual mediums. But this
fact does not undermine the great achievements done by this great
man of science, that ultimately made the first experimental
observations of new phenomena and build the grounds of this new area
of scientific research, known as plasma science. Among his
achievements we can refer the invention of the spinthariscope, a
remarkable particle's detector using a zinc sulfide screen to detect
alpha particles, and the study of radioactive materials. And the
nowadays wonderful toy, the Crookes radiometer, which ultimately
gave one of the first evidence of the existence of atoms and
supported the kinetic theory of gases.

Francis Chen in his renowned book on plasma science advances with a
meaning which contains some weird suggestion:
\begin{quote}
"The word 'plasma' seems to be a misnomer. It comes from the Greek
$\pi \lambda \alpha \sigma \mu \alpha$, -$\alpha \tau o \varsigma$,
$\tau$\'{o}, which means something molded or fabricated. Because of
collective behavior, it often behaves as if it had a mind of its
own"~\cite{Chen_1}.
\end{quote}

However, the use of this term to describe an ionized gas in a
written paper is due to the Nobel prize winning American chemist
Irving Langmuir (1881-1957) in 1928~\cite{Langmuir1}. It seems that
Langmuir was reminded of the way blood plasma carries red and white
corpuscles in a way that somehow similar to an electrified fluid
carrying electrons and ions~\cite{Tonks_1}.

This short note intended to recall the origin of the word and the
pre-logical epistemological viewpoint sustaining at its beginning
the actual branch of plasmas science.

\bibliographystyle{amsplain}
\bibliography{Doc2}

\end{document}